# SAMPLING BASED APPROACHES TO HANDLE IMBALANCES IN NETWORK TRAFFIC DATASET FOR MACHINE LEARNING TECHNIQUES

Raman Singh[1], Harish Kumar[2], and R.K. Singla[2]

[1,2]University Institute of Engineering and Technology, Panjab University, Chandigarh, India
[1]raman.singh@ieee.org, [2]harishk@pu.ac.in
[3]Department of Computer Science and Applications, Panjab University, Chandigarh
[3]rksingla@pu.ac.in

## ABSTRACT

*Network traffic data is huge, varying and imbalanced because various classes are not equally distributed. Machine learning (ML) algorithms for traffic analysis uses the samples from this data to recommend the actions to be taken by the network administrators as well as training. Due to imbalances in dataset, it is difficult to train machine learning algorithms for traffic analysis and these may give biased or false results leading to serious degradation in performance of these algorithms. Various techniques can be applied during sampling to minimize the effect of imbalanced instances. In this paper various sampling techniques have been analysed in order to compare the decrease in variation in imbalances of network traffic datasets sampled for these algorithms. Various parameters like missing classes in samples, probability of sampling of the different instances have been considered for comparison.*

## 1. INTRODUCTION

With growth of networked machines lot of data is transferred among them. Preventive actions against any of the network anomaly can be taken only by analysing this data. But to process whole data is not feasible due to ever growing network traffic. Hence, newer techniques like Machine learning (ML) are used for the analysis. These techniques need traffic samples. Network data may have imbalances which is performance degrader factor for these techniques. Learning from imbalanced dataset may be biased because of unequal distribution of instances. Most of standard ML algorithms assume that instances are equally distributed among classes and hence if dataset is imbalanced the outcomes may be biased and fails to properly represent statistical properties of dataset. This is known as learning problem from imbalanced dataset. Network traffic dataset is also imbalanced dataset and techniques to minimize this effect on ML are required. This paper is divided into six sections. Section I introduces topic, Section 2 describes experiment setup used to capture network traffic, Section 3 describes imbalanced network traffic dataset and its issues; Section 4 describes sampling and pre-processing of dataset, Section 5 discuss the result and Section 6 describes conclusions and future scope.

## 2. EXPERIMENT SETUP

To capture network traffic dataset, a network server is configured in Panjab University Campus-Wide Area Network (PU-CAN). Sub-network of PU-CAN which is used to capture dataset provides network service to three boys' hostels covering approximately 500 users. This network is managed by team of administrators and network engineers. Internet facility is providing through Squid Server to ensure controlled access and surveillance on user's internet usage.

Proper firewall system is configured to stop malwares and unauthorized access and attacks. Dynamic Host Configuration Protocol (DHCP) server is used to dynamically distribute Internet Protocol (IP) address. The official IP addresses range which is assigned to users is 172.16.40.1 to 172.16.43.254. Some IP addresses are kept reserved for servers like 172.16.40.1 for Domain Name Server (DNS), 172.16.40.2 for Squid Server and DHCP Server. The IP address of server which is used to capture network traffic dataset is 172.16.40.11. Figure 1 shows the network diagram of sub-network of PU-CAN

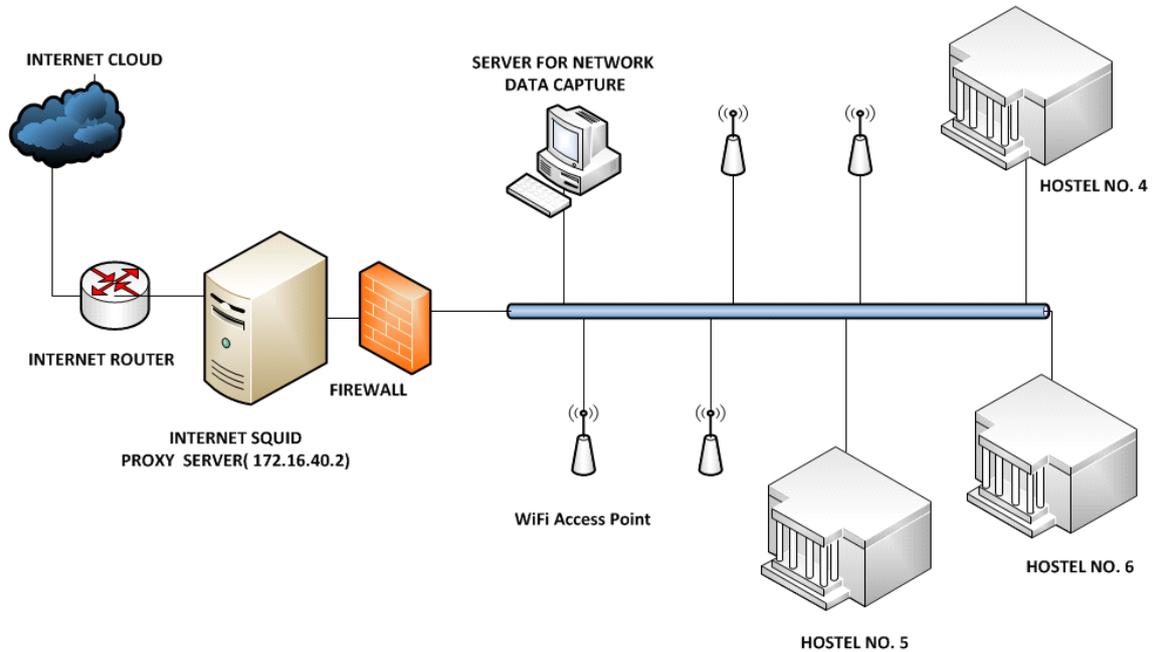

Figure 1. Network Diagram of PU-CAN Used to Capture Traffic Data

## 3. IMBALANCES IN NETWORK TRAFFIC DATASETS

### 3.1. Imbalanced Dataset

At a given point of time on any computer network thousands of packets travel. Capturing each packets and then analysing network traffic dataset in order to detect malwares is very cumbersome job for any Intrusion Detection System (IDS). In network traffic data the instances of malicious packets like malwares, attacks, viruses are very few in number than instances of normal packets. This problem is known as imbalanced dataset problem. This leads to serious problem of under training of the models based on machine learning and subsequently leads to miss-classification. The captured dataset is huge and need to be pre-processed efficiently before any machine learning algorithm is used to analyse it. Imbalanced dataset have unequal distribution of instances among various classes. Some classes may have hundreds and thousands of instances whereas other may have only very few number of instances [1]. In network traffic dataset, one class of packets is present in large numbers while other class has only few instances [2].

### 3.2. Issues in Imbalanced Dataset

Various issues due to imbalanced nature of network traffic dataset are:

i.      Most of ML algorithms used in network traffic classification and profiling to detect malware, take it granted that dataset is balanced in nature & instances/ network packets are equally distributed in all classes, but this is not true in case of network traffic data and results into biased classification [1].

ii.   ML technique will give poor performance on minority classes because distribution of training data may differ from testing data but these techniques are generalized & assume that both dataset have same distribution [3].

iii.   ML algorithms are designed to obtain higher accuracy rate, but this may lead to problem with IDS to detect minority attack patterns. These techniques are guided by standard accuracy rate and it will biased towards covering of majority instances, while on the other hand it is difficult to distinguish between noise & minority instances as both are few [4].

## 4. SAMPLING AND PRE-PROCESSING OF NETWORK TRAFFIC DATASET

Network traffic dataset requires preprocessing steps before machine learning algorithms applied to detect malwares. Preprocessing removes noisy or missing data. Huge network traffic dataset is sampled to increase performance of these algorithms. The various steps involved in preprocessing of network traffic dataset are discussed below:

I) Dataset Generation: The first step is to generate dataset for network traffic classification. Dataset is the divided into training set and test set [5]. Since network traffic dataset is huge, so various sampling techniques are applied to limit size. However characteristics of dataset should not change while using sampling.

II)  Feature Selection and Extraction: Network traffic dataset has various features but all those features may not contribute in classification and intrusion detection. In order to enhance accuracy and performance, important features needs to be selected ignoring other redundant/irrelevant features. New features can also be derived from existing features to increase performance [5].

Commonly available sampling techniques used to sample network traffic dataset are:

a) Random Sampling: is random method to select 'n' instances from population 'P' packets/instances of network traffic dataset [6]. It can be done 'with/without' replacement. Probability of sampling P(s) and size of sampled dataset in percentage of total dataset can be calculated as in equation (1) and (2) below:

*P(s)  = No. of favoured events/ Total no. of events= n/P*                                         (1)

*Size of sampled dataset (%age of total dataset)  = n/P\*100*                                     (2)

b) Systematic Sampling: selects 'n' packets out of total population of 'P' packets by considering a packet after every regular interval starting from a point. For example if dataset size is 10000 packets and 1000 packets needs to be selected then every 10th packets should be selected starting from first instance [7]. Sampling interval can be calculated as in equation (3) below:

*Sampling Interval=Total Population/Packets required=P/n & starting point  =  1st*   (3)
*packet*

c) Stratified Sampling: is capable of discovery of statistical characteristics of network traffic dataset. It is used in heterogeneous dataset where all instances are not of the same type. First, population is divided in to heterogeneous sub-population called strata. Sub-population in each strata is homogenous. Then samples are selected from each strata. While selecting the samples, random/systematic/proportional-to-size sampling can be used [8]. Priory knowledge about characteristics of populations required in stratified sampling. Population 'P' is divided in 'L' groups, and the each group has some different numbers of instances. Then from each group some samples are selected. Total sampled instances n(s) can be calculated by adding sampled instances from each strata as shown in equation (4).

$$n(s) = \sum_{i=0}^{l} n_i,  \quad (4)$$

Probability of sampling for each strata can be calculated by equation (5):

$$P(si) = n_i/P, \text{ where } I = 1 \text{ to } L \quad (5)$$

The size of sampled dataset can be calculated by below given equation (6):

$$\text{Size of sampled dataset( In \%age)} = (\sum_{i=0}^{l} n_i /P), *100 \quad (6)$$

d) Under-Over Sampling: In Under-sampling instances are sampled in fewer rates so that minority and majority classes have equal contribution in sampling method [9]. It is used to balance imbalances in dataset by elimination of instances of majority classes. The drawback is that sometime the loss of information may occur if some particular instances are missed and loss of useful information may occur [10][11]. The instances can be under sampled by some factor. Prior knowledge about dataset is required for deciding this factor. In over-sampling instances of minority classes are synthetically generated up to some pre-defined numbers [10]. Instances are sampled with higher sampling rate. More samples are picked in over-sampling from minority class and few picked in under-sampling from majority class. Drawback is that since the instances are generated synthetically/repeated, redundant information may leads to biased classification. Replicated instances may leads to wrong decisions [11]. Also since network traffic dataset is huge and imbalanced and further if minority classes are large in number then performing over-sampling to reduce imbalances and replicating minority classes instances further increase size of dataset & computational time [10].

## 5. RESULTS AND DISCUSSION

### 5.1. Network Traffic Dataset Characteristics

For analysis of sampling technique on network traffic data, packets are captured and a dataset is prepared. Size this dataset is approximately 16GB comprising of billions of instances. Out of this huge data, a small set of 30000 instances of dataset is taken for testing. Java programming language [12] is used to implement various sampling methods and to analyse results. Dataset used for analysis purpose is named as "Panjab University-Test Data Set" (PU-TDS). It has 30000 instances and 25 different protocols packets. First analysis comprising of number of packets, their percentage and probability with-in sample has been carried out and the results are shown in table 1. Number of instances for each packets associated with different protocol has been calculated. Percentage of packets is also shown. P(s) is probability of packets to be picked in sampling process. Higher the P(s) value, higher the chances of packets to be picked for sampling.

The analysis of test dataset shows that if this dataset is to be classified as per protocols there will be 25 classes associated with each protocol. The probability of packets for each protocol is varying proportional to numbers of packets for these protocols. Some protocols like TCP are present in higher number (11735 instances), while others like HTTP/XML and IAPP has very few presences (only 1 instance). Hence this can be derived that network traffic dataset is imbalanced. This nature of data can mislead the machine learning algorithm, biased classification and miss-classification may occur. Due to huge size of data set, pre-processing steps like sampling are required before feeding it to machine learning algorithms. But probability of certain protocol packets like IAPP or HTTP/XML is minute (0.00003), hence their chances to enter into sample are also rare. It leads to loss of information due to sampling process. Also if machine learning algorithm is used to find malware in intrusion detection system special care should be taken in pre-processing and training as malware instances are very less in number while normal packets may outnumber malicious packets. Due to these characteristics and issues specialize pre-processing/sampling and machine learning techniques should be designed for network traffic classification to detect intrusions/ attacks.

Table 1. Analysis of PU-TDS for probability of sampling and size for each protocol (classes)

| Protocols | No. of Packets | %age of Packets | P(s) |
|---|---|---|---|
| DHCP | 346 | 1.153 | 0.01153 |
| ARP | 3235 | 10.783 | 0.10783 |
| ICMP | 24 | 0.08 | 0.0008 |
| HTTP | 1252 | 4.173 | 0.04173 |
| TCP | 11735 | 39.117 | 0.39117 |
| UDP | 585 | 8.453 | 0.08453 |
| ICMPv6 | 2536 | 5.237 | 0.05237 |
| SSDP | 1571 | 1.95 | 0.0195 |
| NBNS | 642 | 2.14 | 0.0214 |
| MDNS | 118 | 0.393 | 0.00393 |
| LLMNR | 1031 | 3.437 | 0.03437 |
| BROWSER | 193 | 0.643 | 0.00643 |
| TLSv1 | 5669 | 18.897 | 0.18897 |
| DB-LSP-DISC | 75 | 0.25 | 0.0025 |
| DHCPv6 | 462 | 1.54 | 0.0154 |
| DNS | 4 | 0.013 | 0.00013 |
| HTTP/XML | 1 | 0.003 | 0.00003 |
| IAPP | 1 | 0.003 | 0.00003 |
| IGMP | 337 | 1.123 | 0.01123 |
| IPX RIP | 11 | 0.037 | 0.00037 |
| LLC | 146 | 0.487 | 0.00487 |
| NBIPX | 6 | 0.02 | 0.0002 |
| OCSP | 4 | 0.013 | 0.00013 |
| SSL | 13 | 0.043 | 0.00043 |
| XID | 3 | 0.01 | 0.0001 |

### 5.2. Analysis of various sampling techniques using PU-TDS

In order to analyse effect of various sampling techniques on network traffic dataset, various commonly used sampling methods are implemented in Java programming language and tested using PU-TDS.

### 5.2.1. Random Sampling

Packets are randomly selected from PU-TDS. Consider 'n' is the number of packets to be selected for sampling. Experiment is done for different value of 'n' like 500, 1000, 2000, 3000, 5000, 10000, 15000 and 20000. Table 2 shows the percentage of packets selected out of total packets of different protocols for various values of 'n'.

Table 2. Percentage of packets selected for various protocol in random sampling

| n / Protocol | 500 | 1000 | 2000 | 3000 | 5000 | 10000 | 15000 | 20000 |
|---|---|---|---|---|---|---|---|---|
| DHCP | 0.8 | 1.4 | 1.6 | 1.1 | 0.98 | 1.22 | 1.1 | 1.155 |
| ARP | 11.6 | 9.8 | 10.9 | 11.4 | 11.42 | 10.46 | 10.66 | 10.635 |
| ICMP | 0 | 0.2 | 0.05 | 0.067 | 0.1 | 0.12 | 0.08 | 0.1 |
| HTTP | 4.8 | 5.1 | 4.55 | 4.433 | 3.92 | 4.08 | 4.353 | 4.41 |
| TCP | 36.8 | 38.8 | 39.5 | 39.1 | 39.26 | 39.7 | 39.36 | 38.96 |
| UDP | 2.4 | 2.2 | 2.1 | 1.833 | 2.08 | 1.99 | 1.793 | 2.045 |
| ICMPv6 | 8.8 | 9.6 | 8.6 | 9.233 | 8.38 | 8.67 | 8.587 | 8.39 |
| SSDP | 5 | 5.5 | 3.95 | 4.933 | 4.74 | 5.32 | 5.133 | 5.17 |
| NBNS | 2.6 | 2.8 | 1.95 | 2.1 | 2.4 | 2.35 | 2.287 | 2.03 |
| MDNS | 0.4 | 0.3 | 0.25 | 0.167 | 0.42 | 0.4 | 0.333 | 0.43 |
| LLMNR | 2 | 4.1 | 3.9 | 3.7 | 3.54 | 3.48 | 3.373 | 3.48 |
| BROWSER | 0.8 | 0.4 | 0.85 | 0.8 | 0.62 | 0.62 | 0.653 | 0.76 |
| TLSv1 | 19.4 | 15.8 | 18.55 | 17.933 | 18.7 | 18 | 18.873 | 18.965 |
| DB-LSP-DISC | 0.4 | 0.4 | 0.15 | 0.167 | 0.22 | 0.23 | 0.173 | 0.3 |
| DHCPv6 | 1.6 | 1.8 | 1.4 | 1.133 | 1.58 | 1.74 | 1.527 | 1.445 |
| DNS | 0 | 0 | 0 | 0 | 0.02 | 0 | 0.02 | 0.01 |
| HTTP/XML | 0 | 0 | 0 | 0 | 0 | 0 | 0.007 | 0 |
| IAPP | 0 | 0 | 0 | 0 | 0 | 0.01 | 0 | 0 |
| IGMP | 1.8 | 1.1 | 1.05 | 1.367 | 1.12 | 1.03 | 1.16 | 1.1 |
| IPX RIP | 0 | 0.7 | 0.05 | 0.067 | 0.08 | 0.03 | 0.02 | 0.08 |
| LLC | 0.8 | 0 | 0.55 | 0.4 | 0.4 | 0.41 | 0.46 | 0.415 |
| NBIPX | 0 | 0 | 0 | 0.067 | 0 | 0.02 | 0.007 | 0.02 |
| OCSP | 0 | 0 | 0 | 0 | 0 | 0.01 | 0 | 0.015 |
| SSL | 0 | 0 | 0.05 | 0 | 0.02 | 0.08 | 0.033 | 0.075 |
| XID | 0 | 0 | 0 | 0 | 0 | 0.03 | 0.007 | 0.01 |

As the network traffic dataset is imbalanced and heterogeneous, the number of packets selected is varying for different protocols. For some protocols packets the percentage of packets selected are as high as 19.88 while some packets of protocols class are missed by this sampling. It causes loss of information and machine learning will not get proper training on these sampled datasets. Miss-classification may occur. In intrusion detection technique this missing samples may cause harm to network as malicious packets which are fewer in number may not be selected in sampling and Intrusion Detection System (IDS) system may fail to detect attacks. So, this sampling should not be used in sampling of network traffic data. Figure 2 shows the number of classes missed by random sampling as per number of packets selected for sampling.

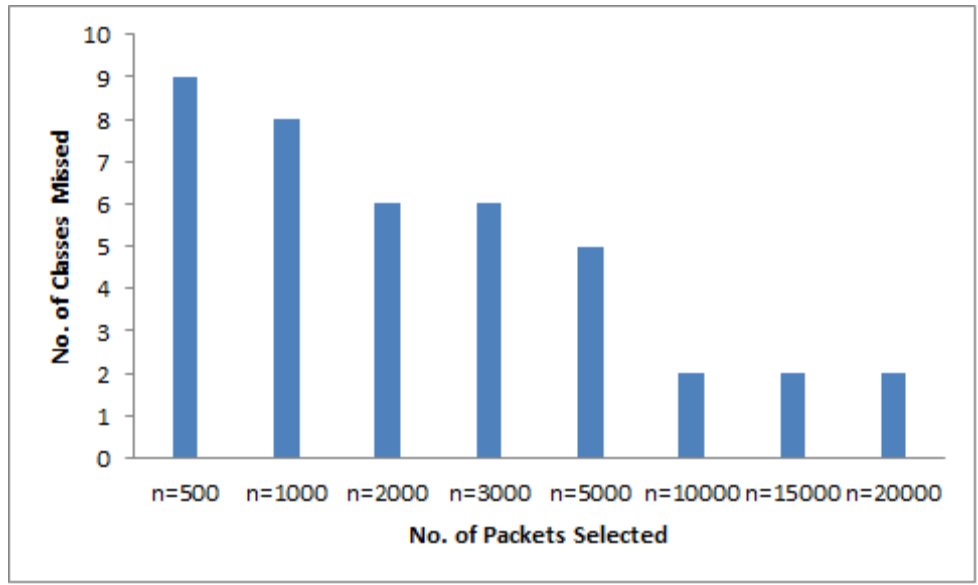

Figure 2. Loss of information in random sampling

From figure-2 it is clear that with increase in sampling factor, there is decrease in number of missing classes. Since packets are randomly selected, it is not necessary that decrease should be uniform. Due to this reason, there are certain exception at n=2000 & n=10000. If sampling factor is 500 (i.e. n=500), then 9 classes out of total 25 classes are missed in sampling and will not contribute in decision making. Further if we increase sampling factor loss of information may decrease but cannot be ruled out completely.

### 5.2.2. Systematic Sampling

Packets are systematically selected from PU-TDS. If every $I^{th}$ packet is selected for sampling, then 'I' is known as sampling factor. As the value of 'I' increases, number of packets selected decreases. In this experiment each 5th, 6th, 7th, 8th, 9th and 10th packet are selected and 6 different sampled dataset are prepared. Table 3 shows the value of 'I', total no. of packets selected for each class and their percentage.

Table 3. Percentage of packets selected for various protocol in systematic sampling

| Protocols | I=5, n=6000 | I=6, n=5000 | I = 7, n=4286 | I = 8, n=3750 | I = 9, n=3334 | I = 10, n=3000 |
|---|---|---|---|---|---|---|
| DHCP | 1.217 | 1.32 | 1.143 | 1.2 | 1.199 | 1.6 |
| ARP | 10.75 | 10.8 | 11.269 | 10.346 | 10.708 | 10.6 |
| ICMP | 0.05 | 0.08 | 0.139 | 0.106 | 0.089 | 0 |
| HTTP | 4.317 | 4.06 | 4.013 | 3.733 | 4.499 | 4.2 |
| TCP | 38.6 | 39.68 | 39.057 | 39.947 | 38.962 | 39.567 |
| UDP | 1.917 | 1.74 | 1.796 | 1.947 | 1.829 | 1.967 |
| ICMPv6 | 7.917 | 8.22 | 8.772 | 8.267 | 8.248 | 7.667 |
| SSDP | 5.8 | 5.3 | 5.016 | 5.28 | 5.579 | 5.867 |
| NBNS | 2.317 | 2.12 | 2.496 | 2.107 | 2.159 | 2.134 |
| MDNS | 0.417 | 0.42 | 0.373 | 0.427 | 0.509 | 0.367 |
| LLMNR | 3.567 | 3.58 | 3.103 | 3.547 | 3.149 | 3.767 |
| BROWSER | 0.683 | 0.54 | 0.607 | 0.533 | 0.569 | 0.767 |
| TLSv1 | 19.083 | 18.52 | 18.992 | 18.64 | 18.866 | 18.434 |
| DB-LSP-DISC | 0.267 | 0.22 | 0.303 | 0.293 | 0.209 | 0.267 |
| DHCPv6 | 1.633 | 1.68 | 1.306 | 1.653 | 1.619 | 1.467 |
| DNS | 0 | 0 | 0 | 0 | 0.029 | 0 |
| HTTP/XML | 0 | 0 | 0 | 0 | 0 | 0 |
| IAPP | 0.017 | 0.02 | 0 | 0 | 0.029 | 0.033 |
| IGMP | 0.95 | 1.06 | 0.979 | 1.36 | 0.959 | 0.8 |
| IPX RIP | 0.033 | 0.08 | 0.069 | 0.027 | 0 | 0 |
| LLC | 0.433 | 0.52 | 0.466 | 0.533 | 0.689 | 0.467 |
| NBIPX | 0.016 | 0.02 | 0.023 | 0.027 | 0 | 0 |
| OCSP | 0 | 0 | 0 | 0 | 0 | 0 |
| SSL | 0 | 0 | 0.069 | 0.027 | 0.059 | 0 |
| XID | 0.017 | 0.02 | 0 | 0 | 0.029 | 0.033 |

From table 3, it can be analysed that some classes are present in majority while others are in minority. Experiment shows that this may miss some classes and can cause loss of information. Figure 3 shows analysis of loss of information in systematic sampling as per different sampling factor values. Number of classes missed is also shown.

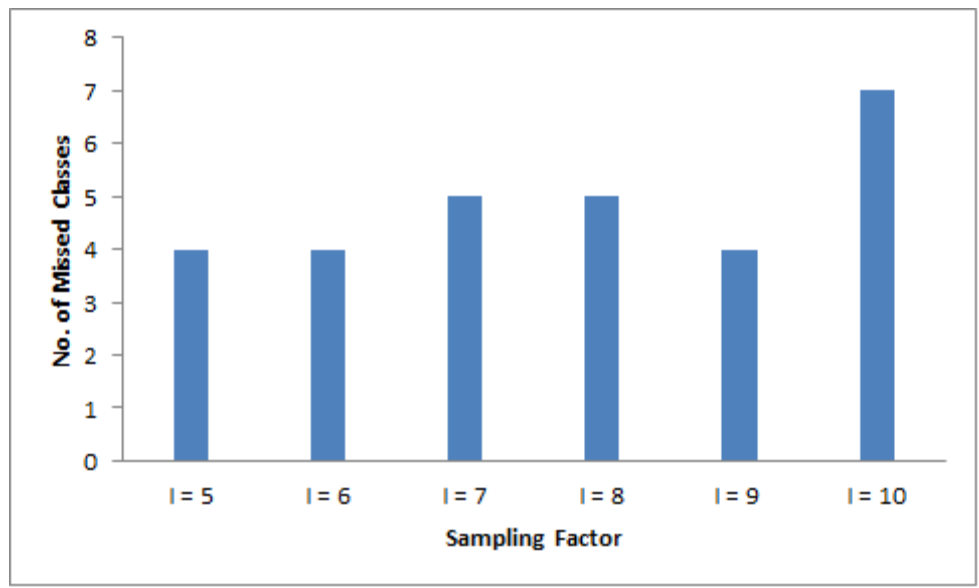

Figure 3. Loss of information in systematic sampling

### 5.2.3. Stratified Sampling

In this sampling different heterogeneous stratas are defined which further contains homogeneous packets. In experiment, 25 different classes are considered depending on protocols. 25 different stratas are defined for each protocol or class. Then for each stratas/classes, 6 experiment performed using systematics sampling and taking values of sampling factor (I) as 5, 6, 7, 8, 9 and 10. The advantage of this sampling is that each of heterogeneous packets will contribute in decision making. Out of each stratas, systematics sampling is used to select packets since some stratas have large number of packets. It is also known as two phase sampling. First each packets is divided into different 25 stratas based on protocol used and then out of each stratas some packets are selected for creating of final sampled dataset. Table 4, shows the results of strata sampling.

Table 4. Percentage of packets selected for various protocol in strata sampling

| Protocols | I = 5, n=6012 | I = 6, n=5010 | I = 7, n=4297 | I = 8, n=3763 | I = 9, n=3347 | I = 10, n=3015 |
|---|---|---|---|---|---|---|
| DHCP | 1.164 | 1.158 | 1.164 | 1.169 | 1.165 | 1.161 |
| ARP | 10.762 | 10.778 | 10.775 | 10.763 | 10.756 | 10.746 |
| ICMP | 0.083 | 0.08 | 0.093 | 0.08 | 0.09 | 0.1 |
| HTTP | 4.175 | 4.172 | 4.166 | 4.172 | 4.183 | 4.179 |
| TCP | 39.039 | 39.042 | 39.027 | 38.985 | 38.96 | 38.939 |
| UDP | 1.946 | 1.956 | 1.955 | 8.424 | 1.942 | 1.957 |
| ICMPv6 | 8.45 | 8.443 | 8.448 | 1.967 | 8.425 | 8.425 |
| SSDP | 5.24 | 5.23 | 5.236 | 5.235 | 5.229 | 5.24 |
| NBNS | 2.146 | 2.136 | 2.141 | 2.153 | 2.151 | 2.156 |
| MDNS | 0.399 | 0.399 | 0.396 | 0.399 | 0.418 | 0.398 |
| LLMNR | 3.443 | 3.433 | 3.444 | 3.428 | 3.436 | 3.449 |
| BROWSER | 0.649 | 0.659 | 0.652 | 0.664 | 0.657 | 0.663 |
| TLSv1 | 18.862 | 18.862 | 18.85 | 18.841 | 18.823 | 18.806 |
| DB-LSP-DISC | 0.25 | 0.259 | 0.256 | 0.266 | 0.269 | 0.265 |
| DHCPv6 | 1.547 | 1.537 | 1.536 | 1.541 | 1.554 | 1.559 |
| DNS | 0.017 | 0.02 | 0.023 | 0.027 | 0.03 | 0.033 |
| HTTP/XML | 0.017 | 0.02 | 0.023 | 0.027 | 0.03 | 0.033 |
| IAPP | 0.017 | 0.02 | 0.023 | 0.027 | 0.03 | 0.033 |
| IGMP | 1.131 | 1.138 | 1.14 | 1.143 | 1.135 | 1.128 |
| IPX RIP | 0.05 | 0.04 | 0.047 | 0.053 | 0.06 | 0.066 |
| LLC | 0.499 | 0.499 | 0.489 | 0.505 | 0.508 | 0.498 |
| NBIPX | 0.033 | 0.02 | 0.023 | 0.027 | 0.03 | 0.033 |
| OCSP | 0.017 | 0.02 | 0.023 | 0.027 | 0.03 | 0.033 |
| SSL | 0.05 | 0.06 | 0.047 | 0.053 | 0.06 | 0.066 |
| XID | 0.017 | 0.02 | 0.023 | 0.027 | 0.03 | 0.033 |

From table 4, it can be analysed that since each heterogeneous packet is considered for creating strata, loss of information is minimal. However, since each stratas have homogeneous packets and number of packets in one stratas may outnumbers packets in others, so imbalances may present.

### 5.2.4. Under-Over Sampling

It is used to balance the ratio of majority and minority classes. Over-sampling and under-sampling is used to correct imbalance of majority & minority class. If packets/instances in one class are very high then up to a fixed number of packets is selected. Otherwise synthetically generate/copied packets are selected. Table 5 shows analysis of re-sampling. Heterogeneous packets are divided into distinguished stratas and then from each strata 'k' number of packets are selected randomly. If number of packets in strata/class is greater than 'k', then randomly 'k' packets are selected. If it is less than 'k', then 'k' packets are randomly copied up to 'k' numbers. Total number of packets selected is 'k' multiply by number of stratas. Values of 'k' taken are 100,200,300,400,500, and 700. Total number of selected packets can be calculated as in equation (7), where 's' is number of stratas.

*Total number of packets selected 'n' = k*s*                                                                                   (7)

Table 1. Percentage of packets selected for various protocol in under-over sampling

| Protocols | k=100, n=2500 | k = 200, n = 5000 | k = 300, n = 7500 | k=400, n=1000 | k=500, n=12500 | k=700, n=17500 |
|---|---|---|---|---|---|---|
| DHCP | 4 | 4 | 4 | 4 | 4 | 4 |
| ARP | 4 | 4 | 4 | 4 | 4 | 4 |
| ICMP | 4 | 4 | 4 | 4 | 4 | 4 |
| HTTP | 4 | 4 | 4 | 4 | 4 | 4 |
| TCP | 4 | 4 | 4 | 4 | 4 | 4 |
| UDP | 4 | 4 | 4 | 4 | 4 | 4 |
| ICMPv6 | 4 | 4 | 4 | 4 | 4 | 4 |
| SSDP | 4 | 4 | 4 | 4 | 4 | 4 |
| NBNS | 4 | 4 | 4 | 4 | 4 | 4 |
| MDNS | 4 | 4 | 4 | 4 | 4 | 4 |
| LLMNR | 4 | 4 | 4 | 4 | 4 | 4 |
| BROWSER | 4 | 4 | 4 | 4 | 4 | 4 |
| TLSv1 | 4 | 4 | 4 | 4 | 4 | 4 |
| DB-LSP-DISC | 4 | 4 | 4 | 4 | 4 | 4 |
| DHCPv6 | 4 | 4 | 4 | 4 | 4 | 4 |
| DNS | 4 | 4 | 4 | 4 | 4 | 4 |
| HTTP/XML | 4 | 4 | 4 | 4 | 4 | 4 |
| IAPP | 4 | 4 | 4 | 4 | 4 | 4 |
| IGMP | 4 | 4 | 4 | 4 | 4 | 4 |
| IPX RIP | 4 | 4 | 4 | 4 | 4 | 4 |
| LLC | 4 | 4 | 4 | 4 | 4 | 4 |
| NBIPX | 4 | 4 | 4 | 4 | 4 | 4 |
| OCSP | 4 | 4 | 4 | 4 | 4 | 4 |
| SSL | 4 | 4 | 4 | 4 | 4 | 4 |
| XID | 4 | 4 | 4 | 4 | 4 | 4 |

From table 5, it can be analysed that imbalances of data can be corrected by increasing instances of minority classes and decreasing that of majority classes. Then each class will have fair and equal contribution in decision making. In IDS, it can be used to detect malwares as these packets are few. If instance of normal packets and malicious packets made balanced then intrusions and attacks on networks can be detected easily by using ML algorithms. Disadvantage is that since imbalances are corrected in network traffic dataset by reducing majority classes and increasing minority classes, sometimes ML algorithms get wrong training by considering synthetically created minority instances as repeated attack patterns. This should be handled properly by suggesting new and improved ML techniques specifically for IDS.

## 6. CONCLUSIONS AND FUTURE SCOPE

Issues of imbalances in network traffic dataset are discussed in this paper. Since this dataset is very large, sampling should be used to increase performance before machine learning algorithm should be employed on dataset for intrusion detection system. Also if sampling is performed on imbalanced network traffic dataset loss of information may happen and IDS may give biased or wrong results. Various commonly available sampling methods are discussed and experiments are performed to check worthiness of these sampling techniques. Dataset is collected using PU-CAN and named as PU-TDS. In random and systematic sampling loss of information may occur which may cause wrong decision making, since malicious packets may be missed during sampling and IDS may fail to detect malwares and attacks. Strata sampling can be used to overcome problem of loss of information as in strata sampling minimum one packet of each heterogeneous packet will be selected but strata sampling does not overcome the imbalances of network traffic dataset. Under-sampling & over-sampling may be used to overcome problem of imbalances in network traffic dataset and each heterogeneous class including normal and malicious will have equal contribution of decision making and malware will be detected efficiently. Problem with under-over sampling is that synthetically generated and repeated minority class packets may seems to be wrong patterns of attacks. To overcome this issue new and improved sampling approaches should be worked on for network traffic dataset to efficiently handle imbalances.